\documentclass[%
 reprint,
 amsmath,amssymb,
 aps,
]{revtex4-1}

\usepackage[utf8]{inputenc}
\usepackage{graphicx}
\usepackage{hyperref}
\hypersetup{colorlinks = true, linkcolor = [rgb]{0.19411,0.51882,0.667058}, urlcolor = [rgb]{0.125490,0.29542,0.1647058}, citecolor = [rgb]{0.75882,0.37411,0.14117}}
\usepackage{amsmath}
\usepackage{blkarray}
\usepackage{amsfonts}
\usepackage{amssymb}
\usepackage{siunitx}
\usepackage{braket}
\usepackage{tikz}
\usepackage{import}
\usepackage{dsfont}
\usepackage{color,soul}

\usepackage{changes}

\newcommand*{\I}{\mathrm{i}\mkern1mu}

\begin{document}

\title{Spin-augmented observables for efficient photonic quantum error correction}

\author{Elena Callus}\email{ecallus1@sheffield.ac.uk}
\author{Pieter Kok}\email{p.kok@sheffield.ac.uk}
\affiliation{Department of Physics and Astronomy, The University of Sheffield, Sheffield, S3 7RH, UK}

\begin{abstract}\noindent

We demonstrate that the spin state of solid-state emitters inside micropillar cavities can serve as measure qubits in syndrome measurements. The photons, acting as data qubits, interact with the spin state in the microcavity and the total state of the system evolves conditionally due to the resulting circular birefringence. By performing a quantum non-demolition measurement on the spin state, the syndrome of the optical state can be obtained. Furthermore, due to the symmetry of the interaction, we can alternatively choose to employ the optical states as measure qubits. This protocol can be adapted to various resource requirements, including spectral discrepancies between the data qubits and codes with modified connectivities, by considering entangled measure qubits. Finally, we show that spin-systems with dissimilar characteristic energies can still be entangled with high levels of fidelity and tolerance to cavity losses in the strong coupling regime.
\end{abstract}

\date{\today}

\maketitle

Linear optical quantum computing with single photons becomes resource-inefficient and requires a high overhead due to weak photon--photon interactions, making multi-qubit gates difficult to implement \cite{Kok2007}. However, this drawback can be overcome at the measurement stage if one can resolve amongst a broader class of observables, e.g., performing measurements of two-photon qubits such as Bell states \cite{Knill2001}. This would thereby allow for the execution of non-linear gates more efficiently. Although quantum dot (QD) spin systems tend to be too short-lived for useable long-term memories, they interact efficiently with light. The spin--photon interaction augments photonic quantum information processing, with important applications in photonic state measurements. This complements linear optical quantum computing and can dramatically increase its efficiency. Here we propose an application of this interaction to the measurement of a larger class of qubit observables.

\begin{figure}[b]
\includegraphics[width=0.9\columnwidth]{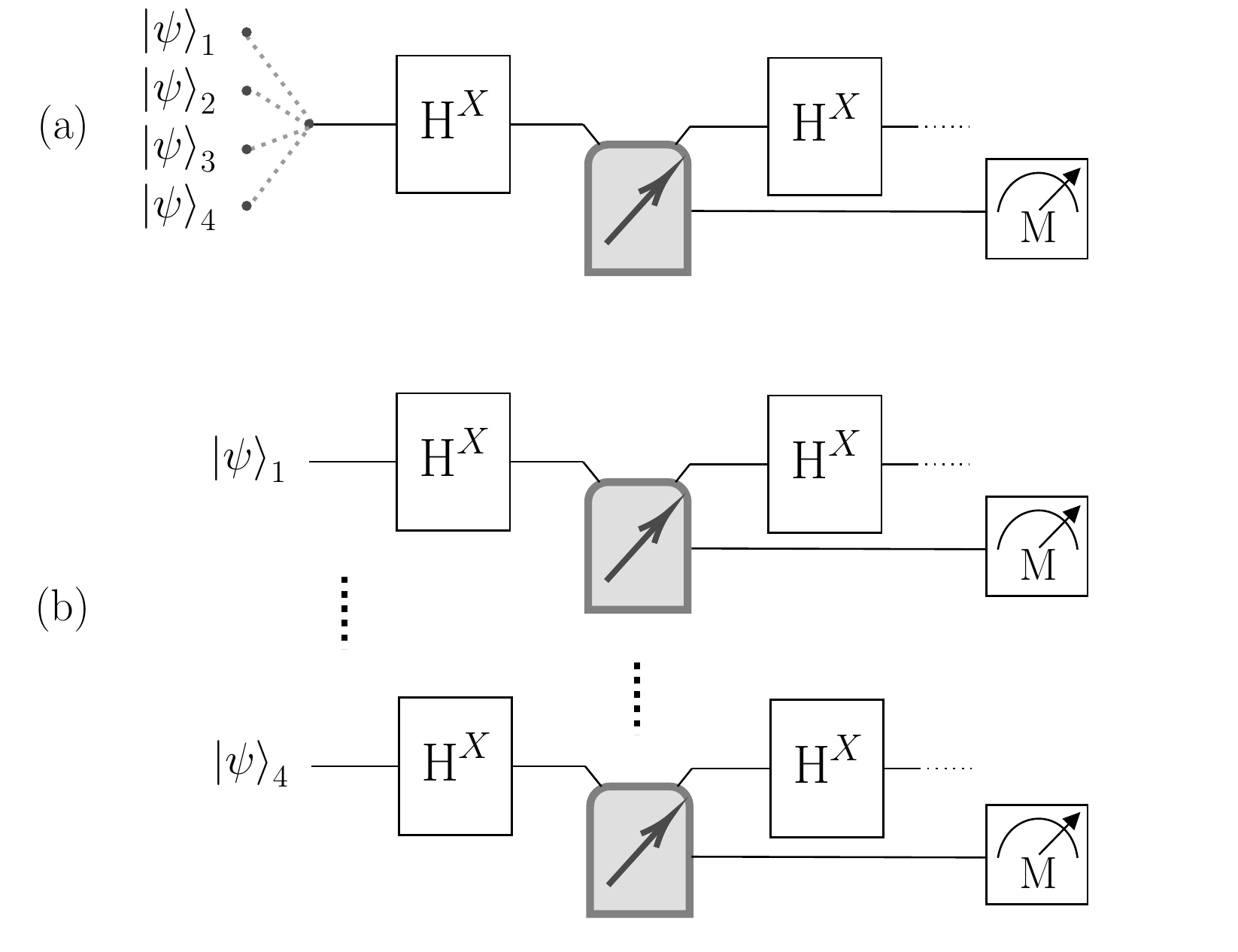}
\caption{Schematic of the stabilizer measurement setup, with (a) a single QD and (b) multiple entangled QDs. The optical states, $\ket{\psi}$, interact with the spin state(s) either (a) successively or (b) in parallel. A spin measurement, M, in the $\hat{X}$ basis is performed as a final step. Hadamard gates, H$^X$, pre- and post-interaction are applied only in the case of a star measurement, $\mathsf{X}_s$.}
\label{fig:setup}
\end{figure}

\begin{figure}[]
\includegraphics[width=1.0\columnwidth]{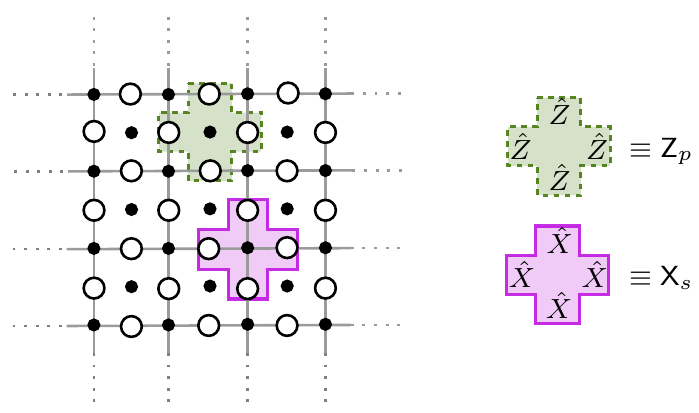}
\caption{A schematic of the two-dimensional array of data (open circles) and measure (black circles) qubits of a surface code. The plaquette, $\mathsf{Z}_p$, and star, $\mathsf{X}_s$, measurement operators are shown in green (dotted outline) and pink (solid outline), respectively.}
\label{fig:sandp}
\end{figure}

The spin--photon interface is a promising candidate for applications in quantum information technologies and quantum communication \cite{Kimble2008,Atatre2018}. The low decoherence rate of the photon renders it suitable as a flying qubit, transporting information over large distances and interacting readily with the solid-state spin, which acts as a stationary qubit. Over the last few years, various systems belonging to this family have been extensively studied with the aim of applications in various quantum technologies, yielding the development of, e.g., photonic quantum gates \cite{Hacker.2016,Duan.2004} and optical non-linearities \cite{Javadi.2018,Javadi.2015}, as well as entanglement of remote spin states \cite{Delteil.2015,Young2013,Cirac.1997}, photon polarisation \cite{Hu.2008a} and spin--photon states \cite{Economou.2016}. The circular birefringence arising from the optical selection rules for a spin state confined in a cavity \cite{Young2011} has been used to develop schemes for, e.g., quantum teleportation \cite{Hu2011}, quantum non-demolition measurements \cite{Hu2009} and entanglement beam splitters \cite{Hu2009a}. Furthermore, this system also has applications in the design of complete and deterministic Bell-state analyzers \cite{Bonato.2010,Hu2011}, a marked improvement over what is possible using just linear optics \cite{Calsamiglia2001}. Here, the spin--photon system measures the qubit parity whilst information about the symmetry is obtained using linear optics. 

In this work, we will discuss the application of spin--photon interfaces to carry out efficient photonic stabilizer measurements in the surface code. A key objective in quantum physics is the physical realisation of fault-tolerant quantum computers, necessitating the development of quantum error detection and correction \cite{Shor1995}. Surface codes are a well-studied set of stabilizer codes designed for the implementation of error-corrected quantum computing \cite{Roffe.2019}. The first proposal was presented by Kitaev in the form of the toric code \cite{Kitaev.2003,Kitaev.1997}, assuming periodic boundary conditions that allow it to be mapped onto a torus. This was later generalised to planar versions with different variations in the boundary conditions \cite{Bravyi.1998,Freedman.1998,Fowler.2012}. The surface code considers a 2D square lattice arrangement of data and measure qubits, with the latter being used to detect errors and perform stabilizer measurements of the encoded data qubits.

Stabilizer measurement is one type of error detection technique, indicating the presence of possible noisy errors in the physical data qubits \cite{Nielsen2012,Gottesman1997}. It consists of a series of projective measurements performed on specific sets of qubits, with the measurement outcomes, or syndromes, indicating the location and type of error. Given that a direct measurement of the physical data qubits interferes with the coherence of the state and destroys the encoded information, the measurements are performed on entangled measure qubits. There exist several approaches when it comes to the physical implementation of quantum error correction, with platforms including photonic architectures \cite{Bell2014,Yao2012,Aoki2009}, superconducting circuits \cite{Andersen2020,Rist2015,Kelly2015}, trapped atomic ions \cite{Linke2017,Lanyon2013} and nitrogen-vacancy centres \cite{Cramer2016} having been experimentally explored. 

Using solid-state QDs trapped inside micropillar cavities and scattering interactions at the single-photon level, the total state of the optical and spin sub-systems evolves conditionally, with entanglement occuring in the presence of coherent errors. This allows us to perform a quantum non-demolition measurement of one of the two subsystems, effectively retrieving information about the state of the other. Letting the optical and spin states serve as data and measure qubits, we show this interaction and measurement process can be used to extract the syndrome, as shown schematically in Fig. \ref{fig:setup}. The scheme also has the advantage that the assignment of the data and measure qubits can be swapped around. Moreover, we will discuss the use of multiple entangled measure qubits as a means of reducing resource requirements, accommodating for possible spectral variations between the data qubits and codes that consider various connectivities between the qubits.

The detection of errors in the data qubits is performed by means of syndrome measurements. The measurement operators, or stabilizers, for a surface code are comprised of star, $\mathsf{X}_s =\prod_{j\in \text{star}(s)}\hat{X}_j $, and plaquette operators, $\mathsf{Z}_p=\prod_{j\in \text{plaq}(p)}\hat{Z}_j$, where $\hat{X}$ and $\hat{Z}$ are the Pauli-X and Pauli-Z operators. The operators act on either the four data qubits that are adjacent to a vertex, said to belong to a star $s$, or adjacent to a face, said to belong to a plaquette $p$, as shown in Fig. \ref{fig:sandp}. Furthermore, the operators all commute, with $[\mathsf{X}_s,\mathsf{X}_{s'}]=[\mathsf{Z}_p,\mathsf{Z}_{p'}]=[\mathsf{X}_s,\mathsf{Z}_{p}]=0$. The eigenstates of $\hat{Z}$ are $\ket{0}$ and $\ket{1}$, and those of $\hat{X}$ are $\ket{\pm}\propto\left(\ket{0}\pm\ket{1}\right)$. The graph of the surface code is stabilized by $\mathsf{X}_s$ and $\mathsf{Z}_p$. The eigenvalues obtained from the measurement of these operators indicates the possible presence of errors, and depends on the parity of the state, with an eigenvalue of $+1$ ($-1$) corresponding to a state with even (odd) parity. The state of the data qubits may be initialised such that they are simultaneous eigenstates of all the stabilizer operators with eigenvalues of $\pm1$, referred to as the quiescent state. The standard method of extracting the qubit syndrome involves the implementation of a CNOT gate on each of the data qubits belonging to a star or plaquette, with the measure qubit serving as the control.

\begin{figure}[]
\includegraphics[width=0.8\columnwidth]{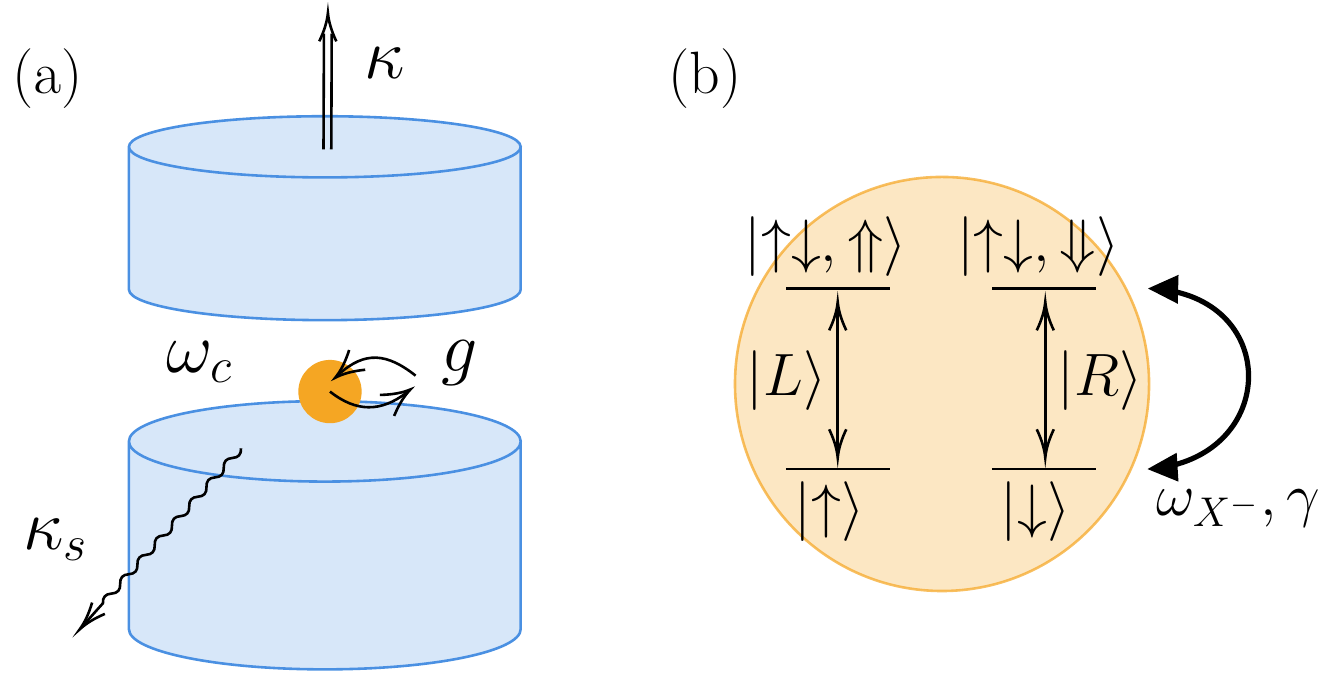}
\caption{(a) A micropillar cavity, with resonance frequency $\omega_c$, coupled to a QD with strength $g$. The field in the cavity couples to the output mode and lossy modes with rates $\kappa$ and $\kappa_s$, respectively. (b) The polarisation- and spin-dependent coupling rules for the QD with central frequency $\omega_{X^-}$ and decay rate $\gamma$.}
\label{fig:micropillar}
\end{figure}

The physical spin setup, demonstrated in Fig. \ref{fig:micropillar}, consists of a single-sided micropillar with two distributed Bragg reflectors on both ends, where only one side is fully reflective and the other end is partially transmissive. The cavity mode couples to an electron spin in the form of a charged QD contained within the micropillar cavity. Given the optical selection rules, the interaction of a photon within the cavity becomes polarisation and spin dependent \cite{Warburton2013}. In the case of a negatively charged QD, the $\ket{\uparrow}$ ($\ket{\downarrow}$) spin state can be optically excited to the negative trion, $X^-$, state $\ket{\uparrow\downarrow,\Uparrow}$ ($\ket{\uparrow\downarrow,\Downarrow}$) by absorption of a left-handed (right-handed) circularly polarised photon, $\ket{L}$ ($\ket{R}$). Cross-transitions between the lower and higher energy states are not allowed by the conservation of angular momentum.

The post-interaction reflection coefficient for a single-sided cavity coupled to a QD is given by \cite{Hu.2008}
\begin{equation}\label{eq:rh}
r_h(\omega)=\frac{\left[\I\left(\omega_{X^-}-\omega\right)+\frac{\gamma}{2}\right]\left[\I\left(\omega_{c}-\omega\right)+\frac{\kappa_s}{2}-\frac{\kappa}{2}\right]+g^2}{\left[\I\left(\omega_{X^-}-\omega\right)+\frac{\gamma}{2}\right]\left[\I\left(\omega_{c}-\omega\right)+\frac{\kappa_s}{2}+\frac{\kappa}{2}\right]+g^2},
\end{equation}
where $\omega$, $\omega_c$ and $\omega_{X^-}$ represent the frequencies of the photon, the cavity mode and the trion transition, respectively; $\gamma$ represents the decay rate of the $X^-$ dipole, $\kappa$ and $\kappa_s$ are the cavity decay rates into the output and the lossy side modes, respectively, and $g$ is the coupling strength between the QD and the cavity field. When the photon does not couple to the QD due to the selection rules, the only contribution to the reflection coefficient is from the (empty) cavity interaction. Setting $g=0$, we can characterise the cold cavity interaction by
\begin{equation}
r_0(\omega)=\frac{\I\left(\omega_{c}-\omega\right)+\frac{\kappa_s}{2}-\frac{\kappa}{2}}{\I\left(\omega_{c}-\omega\right)+\frac{\kappa_s}{2}+\frac{\kappa}{2}}.
\end{equation}
We will consider only the resonant interaction case in this work, where $\omega_c=\omega_{X^-}$,  and allow detuning of the photon frequency, $\omega$, where $\delta=\omega_c-\omega=\omega_{X^-}-\omega$. For small enough cavity losses $\kappa_s$, one sees that $|r_0(\omega)|\simeq 1$ for all frequency detuning, whilst $|r_h(\omega)|\simeq 1$ except in the region of $\delta= \pm g$, when in the strong-coupling regime with $g>\left(\kappa+\kappa_s\right)/4$. We apply a frequency detuning such that the difference in phase shifts imparted during the coupled and the cold cavity interactions is $\pm\pi/2$. This means that the $\delta$ is set such that $\tilde{\phi}(\omega)\equiv\phi_h(\omega)-\phi_0(\omega)=\pm \pi/2$, where $\phi_i(\omega)=\arg\left[r_i(\omega)\right]$ for $i=h,0$. We will drop the notation for frequency dependence for ease of readability.

Fig. \ref{fig:setup} shows a diagrammatic setup for the syndrome measurement, where we first consider situation (a) with four photons interacting sequentially with a single spin. The photons serve as data qubits with $\ket{L}$ and $\ket{R}$ encoding the logical $\ket{0}$ and logical $\ket{1}$ qubit states, respectively, whilst the spin states act as the measure qubits. The Hadamard gates are applied pre- and post-interaction only when performing a star measurement, $\mathsf{X}_s$, such that the $\hat{X}$-basis eigenstates transform as $\ket{+}\leftrightarrow\ket{g}$ and $\ket{-}\leftrightarrow\ket{e}$. The electron spin state is initialised to $\ket{+_S}=\left(\ket{\uparrow}+\ket{\downarrow}\right)/\sqrt{2}$, however we note that the procedure also works for an initial spin state given by $\ket{-_S}=\left(\ket{\uparrow}-\ket{\downarrow}\right)/\sqrt{2}$. Letting the four photons belonging to a plaquette or star set interact with the spin system sequentially in time, and assuming $|r_0 \left(\omega\right)|=|r_h\left(\omega\right)|= 1$, a photonic eigenstate and the electron spin state evolve together, up to a global phase, as
\begin{equation}\label{equation:state}
\begin{split}
\bigotimes_{\substack{j\in \text{star}(s) \\ \text{or }\text{plaq}(p)}}&\ket{i_j}\otimes\ket{+_S}\rightarrow \bigotimes_{\substack{j\in \text{star}(s) \\ \text{or }\text{plaq}(p)}} \exp\left(\I\tilde{\phi}\delta_{i_jL}\right)\ket{i_j}\\
&\otimes\left[\ket{\uparrow}+\prod_{\substack{k\in \text{star}(s) \\ \text{or }\text{plaq}(p)}}\exp\left[-\I\left(\phi_{k,\uparrow}-\phi_{k,\downarrow}\right)\right]\ket{\downarrow}\right]/\sqrt{2},\\
\end{split}
\end{equation}
where $\ket{i_j}\in\{\ket{L},\ket{R}\}$, $j$ and $k$ index the same four photonic qubits in a plaquette or star set, $\delta_{i_j L}$ is the Kronecker delta, and $\phi_{j,\ast}$ is the phase shift resulting from the interaction between the photonic state $\ket{i}_j$ and spin state $\ket{\ast}\in\left\{\ket{\uparrow},\ket{\downarrow}\right\}$. The frequency detuning, $\delta$, is set such that $\tilde{\phi}=\pm\pi/2$, resulting in $\left(\phi_{k,\uparrow}-\phi_{k,\downarrow}\right)=\pm\pi/2$ ($\mp\pi/2$) for a left-handed (right-handed) circularly polarised photon. The relative phase shift between the two spin states accumulates with every spin--photon interaction. The total phase shift imparted from two orthogonally polarised photons is zero, whilst that resulting from pairs of identically polarised photons is $\pm\pi$. Given the set of all eigenstates, the spin state evolves to $\ket{+_S}$ ($\ket{-_S}$) for an even (odd) parity photonic state. By measuring the spin in the $\hat{X}$-basis, we can therefore perform a quantum non-demolition measurement that reveals the syndrome of the data qubits in a complete and efficient manner. The phase shift acquired by the individual photonic eigenstates post-measurement is shown in Eq. \ref{equation:state} and has two contributions. The key contribution to the imparted phase shift is $\exp\left(\I\tilde{\phi}\delta_{i_jL}\right)$, which introduces unwanted phase flips to the encoded state. This is corrected by a polarisation-dependent phase shift acting only on the right-circularly polarised state such that $\ket{R}\rightarrow\exp\left(\I\tilde{\phi}\right)\ket{R}$, possible to achieve in a passive manner using linear optics. This rotation corrects the state, irrespective of the physical state, preserving the original pre-measurement encoded state, including any detected errors. The procedure is similar when we account for the presence of various boundary conditions in the surface code (see SM).

Next, we consider the confidence in the spin read-out \cite{Kok2001} as an appropriate figure of merit for the measurement performance given possible variations in the frequency detuning, $\delta$, from the optimal. The ground state of the planar code is given by \cite{Nielsen2012}
\begin{equation}
\ket{\psi_0} \propto \prod_s\left(\mathds{1}+\mathsf{X}_s\right)\ket{0}^{\otimes n},
\end{equation}
where $n$ is the number of physical data qubits and the product is over the whole set of stars $s$. The state is assumed to be prone to coherent errors that can be modelled by applying a Pauli channel of the form
\begin{equation}
\mathcal{E}\left(\rho\right)=(1-p)\rho+x\hat{X}\rho\hat{X}+y\hat{Y}\rho\hat{Y}+z\hat{Z}\rho\hat{Z}
\end{equation}
to each individual physical qubit, where $x,y,z$ are the probabilities of the respective Pauli errors and $p=x+y+z$ is the physical qubit error rate. Since plaquette (star) operators detect only $X$($Z$)-type errors, we may address the performance of each measurement type individually. For our confidence measure, we may simply assume that both star and plaquette measurements also factor in any possible $Y$-type errors, since $\hat{Y}=\hat{Z}\hat{X}$. We can then express the confidence in a spin read-out of $\ket{\pm}$ by
\begin{equation}
\frac{\text{Tr}\left[\mathbb{P}_\pm\otimes\ket{\pm_S}\bra{\pm_S}\left\{U\left(\mathcal{E}\left(\rho\right)\otimes\ket{+_S}\bra{+_S}\right)U^\dagger\right\}\right]}{\text{Tr}\left[\mathds{1}\otimes\ket{\pm_S}\bra{\pm_S}\left\{U\left(\mathcal{E}\left(\rho\right)\otimes\ket{+_S}\bra{+_S}\right)U^\dagger\right\}\right]},
\end{equation}
where $U$ is the spin- and polarisation-dependent two-qubit gate describing the interaction and $\mathbb{P}_\pm$ is the projection operator onto the $\pm1$ eigenstates.

\begin{figure}[]
\includegraphics[width=\columnwidth]{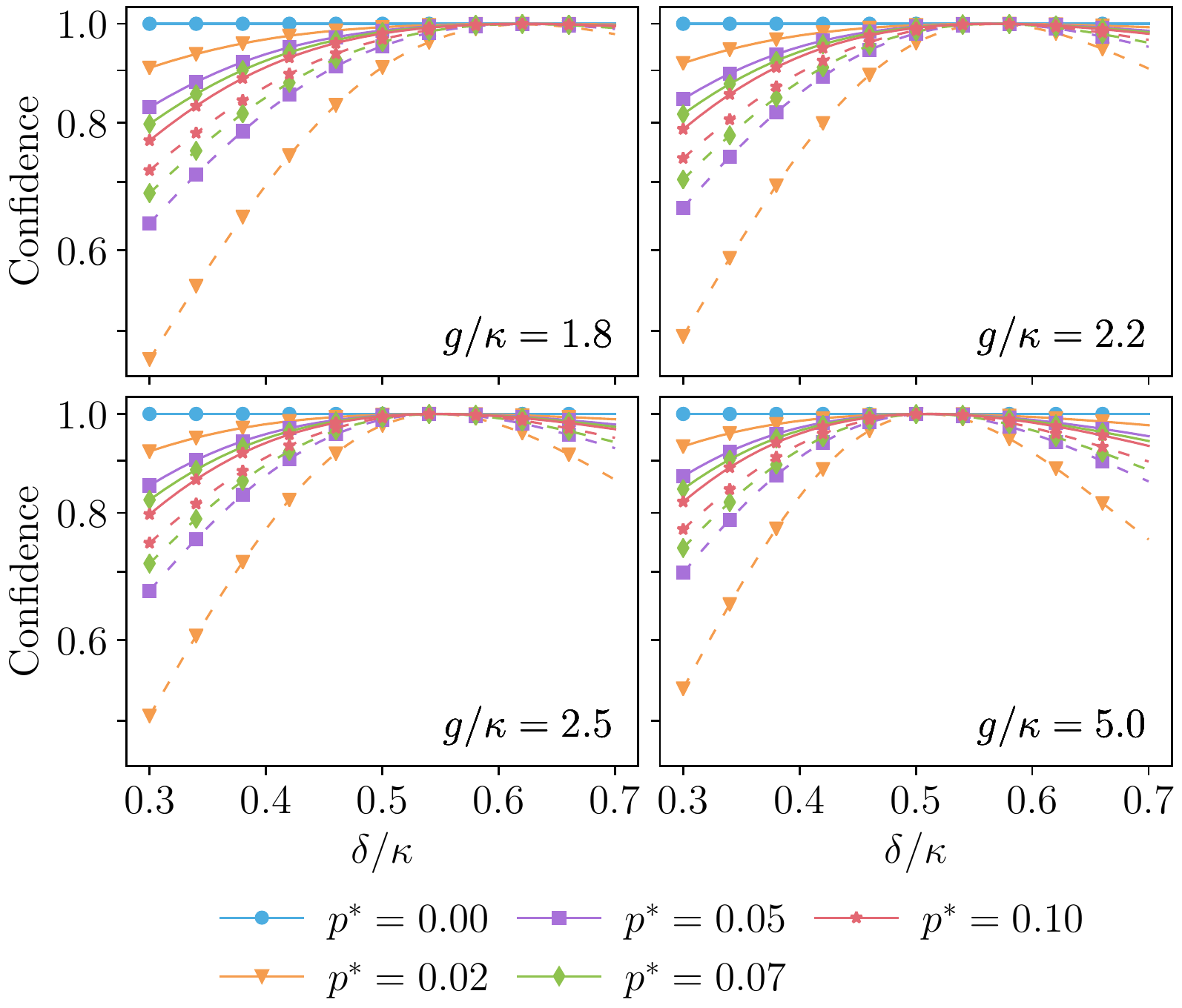}
\caption{Confidence in the $\ket{+_S}$ (solid) and $\ket{-_S}$ (dashed) spin read-outs as a function of the frequency detuning $\delta=\omega_c-\omega=\omega_{X^-}-\omega$ for varying coupling strengths, $g$, and physical qubit error probabilities of either $X$- or $Z$-type, $p^*$. The normalised linewidth, $\gamma/\kappa$, is set to 0.1.}
\label{fig:confidence}
\end{figure}

We show in Fig. \ref{fig:confidence} the confidence in the two types of plaquette and star measurement outcomes for different coupling strengths $g$ and for various physical qubit error probabilities as a function of the frequency detuning $\delta$. We consider here only the strong-coupling regime, as it is in this case that we satisfy the requirement that $|r_h|\rightarrow 1$. Current experimental values show coupling strengths with $g/\kappa$ reaching values of up to around 2.4 \cite{Volz2012,Reitzenstein2007}. We note here that the distance of the code, i.e., the measure of the number of physical qubits used to encode a logical qubit, does not have an effect on the confidence value, and that the plaquette and star measurements show the same type of behaviour due to their equivalence up to a Hadamard gate. We see that the confidence in the $\ket{-}$ spin state measurement tends to be lower. This is due to the way the coefficients accumulate in the presence of errors: the accumulation in such cases builds up in an uneven way, resulting in a more volatile behaviour as the $\delta$ is varied. Furthermore, we show that the proposed scheme is robust and tolerant to deviations in the frequency detuning from the optimal.

Next, we consider a setup where the syndrome measurement is performed in parallel on the incoming photons. This may be done in order to optimise for the type of resources required as well as to accommodate for possible differences in the spectral characteristic of the data qubits. Due to the linear nature of our transformation, the total phase shift is equivalent to the sum of the individual interactions and therefore the syndrome measurement can be done using two or four measure qubits per stabilizer measurement. The spins need to be entangled into a general GHZ-state up to any Pauli operations. Each photon is then allowed to ineract with exactly one of the spins (or, in the case of a two-qubit register, two photons interacting with each spin), such that the interaction satisfies the conditions specified for the single-qubit register. Finally, all the spins are measured in the $\hat{X}$-basis, as was done in the single-qubit register, in order to extract the syndrome.

A setup making use of two or four measure qubits would require fewer photon switches and optical circulators, and would cater for a larger spectral variation between the photons. On the other hand, this calls for entanglement generation, which may require additional resources in terms of time and physical components. One way of entangling the spin states of two QDs is to allow a linearly polarised photon to interact with each state sequentially \cite{Young2013,Hu.2008}. This results in a so-called optical Faraday rotation which rotates the polarisation and, given initial spin states $\left(\ket{\uparrow}+\ket{\downarrow}\right)/\sqrt{2}$ and $\tilde{\phi}=\pm\pi/2$, evolves the spin--photon state to $-\I\ket{V}\left(\ket{\uparrow\uparrow}-\ket{\downarrow\downarrow}\right)\pm\I\ket{H}\left(\ket{\uparrow\downarrow}+\ket{\downarrow\uparrow}\right)$, up to normalisation. Therefore, by measuring the polarisation of the photon, the spin state is projected onto a maximally entangled state. This protocol can be extended to four spin states by entangling another pair and then entangling together one spin state from each pair. 

In the case of photonic data qubits with different frequencies, we would require spectrally different QD-spin systems in order to satisfy the condition of $\tilde{\phi}=\pm\pi/2$. In such cases, the entanglement procedure outlined above may still be used to generate states with high fidelity, albeit the heralded efficiency of the procedure is reduced. By setting the frequency of the linearly polarised photon, say $\ket{H}$, such that $\tilde{\phi_1}=-\tilde{\phi_2}$, where $\tilde{\phi_i}$ is the difference in phase shifts for QD-spin system $i$, the state of the total system post-interaction is
\begin{equation}\label{eq:prob}
\begin{split}
\ket{H}&\left(\ket{\uparrow\uparrow}+\ket{\downarrow\downarrow}\right)+\left(e^{\I\tilde{\phi}}\ket{L}+e^{-\I\tilde{\phi}}\ket{R}\right)\ket{\uparrow\downarrow}\\
&+\left(e^{-\I\tilde{\phi}}\ket{L}+e^{\I\tilde{\phi}}\ket{R}\right)\ket{\downarrow\uparrow},
\end{split}
\end{equation}
up to some global phase and normalisation constant. Upon the detection of an orthogonally polarised photon (in this case $\ket{V}$), the spin states would be projected onto the maximally entangled state $\left(\ket{\uparrow\downarrow}-\ket{\downarrow\uparrow}\right)/\sqrt{2}$. Similarly, one can set the photonic frequency such that $\tilde{\phi_1}=\tilde{\phi_2}$, probabilistically generating the entangled state $\left(\ket{\uparrow\uparrow}+\ket{\downarrow\downarrow}\right)/\sqrt{2}$. The efficiency of the entanglement generation increases with the energy detuning until it peaks at around $40-60\%$ of the maximum possible efficiency. This is because a phase shift that maximises the probability of obtaining an orthogonally polarised photon (i.e. $|\tilde{\phi}|\approx\pi/2$) while satisfying the requirement set in Eq. \ref{eq:prob} is easier to achieve in systems that are sufficiently dissimilar.

One physical limitation that needs to be accounted for is the spin decoherence time, $T_2$, whereby the coherence of the superposition of spin states decays mostly due to interactions with nuclear spins, with experimental values for $T_2$ in the range of several \si{\nano \second} \cite{Androvitsaneas2022,Tran2022,Huang2015}. In the case of a single QD, the fidelity of the spin state would reduce by a factor of $\left(1+\exp\left[-t/T_2\right]\right)/2$, where $t$ is the total time taken for all four photons to interact with the spin, with current lifetime values of exciton photons in micropillars reaching a few hundred \si{\pico\second} \cite{Gins2022,Gins2021,Huber2020}, depending on the detuning between the emitter and the cavity. In the case of $n$ QDs, the fidelity would decay by a factor of $\left(1+\exp\left[-nt/T_2\right]\right)/2$. Since the interaction time $t$ is inversely proportional to the register size of the spins, the reduction in fidelity due to spin decoherence when utilising multiple entangled spin states remains the same. Moreover, although there is a reduction in the measurement confidence and fidelity, the spin dephasing has no detrimental effect on the quiescent state of the data qubits once the syndrome has been extracted.

In conclusion, we have shown how the spin--photon interface may be applied to quantum error detection to perform syndrome measurements, specifically by utilising solid-state emitters inside micropillar cavities and optical circular birefringence. Working in the strong-coupling regime, we have also shown that the scheme is robust over the frequency detuning $\delta$ for coupling strengths $g$ routinely reached in experiment, making this proposed scheme a viable practical candidate. Our analysis has centred around the use of the spin state as the measure qubit, however due to the inherent symmetry of the interaction, it is possible to swap around the assignment of the two types of qubits and perform the syndrome measurement with the photonic state instead. Moreover, it might prove to be useful to use entangled spin states in some implementations as increasing the register of measure qubits in this way also allows for flexibility in the connectivity of the code \cite{Chamberland2020,Chamberland2020a} and accommodates for spectral variations between the data qubits. Such a setup may also prove to be a more resource efficient way of physically realising surface codes tailored to biased noise, where Hadamard transformations are applied to certain Pauli matrices of the star and plaquette operators \cite{Tuckett2018,BonillaAtaides2021,Tiurev2022}. We therefore show that entanglement is still possible for QD-systems with varying characteristic energies. This can be done with high levels of fidelity, albeit with lower generation efficiencies, even in lossy systems when working in the strong coupling regime. 

Potential directions for future work include the extension of the proposed scheme to other measurement families in quantum error correction. Examples of these include the logical Pauli operators, acting on the whole column or row of the qubit array \cite{Bravyi2014}; lattice surgery, which results in logical operations on the encoded qubits by means of splitting or merging the qubit lattice \cite{Horsman2012}; and measurements in higher-dimensional hypergraph product codes \cite{Zeng2019}. Other avenues to explore would be the generalisation of a Bell-state analyser to a scheme that would allow for the observation and discrimination between non-maximally entangled states, as well as the measurement of photons for further versatility. It is evident that the spin--photon interface has potential applications in various aspects of optical quantum information processing, proving it to be a versatile and integral component in the design of quantum technology and vastly improving the performance of linear optical quantum computing.

\begin{acknowledgments}
EC is supported by an EPSRC studentship. PK is supported by the EPSRC Quantum Communications Hub, Grant No. EP/M013472/1. The authors thank Armanda Quintavalle, Joschka Roffe, Ruth Oulton and Andrew Young for valuable comments and discussions. 
\end{acknowledgments}

\bibliography{Papers}{}

\clearpage

\widetext

\begin{center}
\textbf{\large Supplemental Material: Spin-augmented observables for efficient photonic quantum error correction}
\end{center}

\setcounter{equation}{0}
\setcounter{figure}{0}
\setcounter{table}{0}
\setcounter{page}{1}
\makeatletter
\renewcommand{\theequation}{S\arabic{equation}}
\renewcommand{\thefigure}{S\arabic{figure}}
\renewcommand{\bibnumfmt}[1]{[S#1]}
\renewcommand{\citenumfont}[1]{S#1}

\section*{Star and plaqeutte measurements}

We assume that the spin, serving as the measure qubit, is initialised in the state $\ket{+_S}=\left(\ket{\uparrow}+\ket{\downarrow}\right)/\sqrt{2}$. The star and plaquette operators, defined as
\begin{equation}
\mathsf{X}_s =\prod_{j\in \text{star}(s)}\hat{X}_j \qquad \text{ and } \qquad
\mathsf{Z}_p=\prod_{j\in \text{plaq}(p)}\hat{Z}_j,
\end{equation}
respectively, where $\hat{X}$ and $\hat{Z}$ are the Pauli-X and Pauli-Z operators, stabilize the surface code. The eigenbasis for these operators may be expressed as
\begin{equation}\label{eq:eigenbasis}
\left\{\bigotimes_{j\in \text{star}(s)}\ket{i}_j \text{s.t.} \ket{i}=\ket{0} \text{or} \ket{1}\right\} \qquad \text{ and } \qquad \left\{\bigotimes_{j\in \text{plaq}(p)}\ket{i}_j \text{s.t.} \ket{i}=\ket{+} \text{or} \ket{-}\right\},
\end{equation}
respectively, where $\ket{\pm}=\left(\ket{0}\pm\ket{1}\right)/\sqrt{2}$. 

We encode the logical $\ket{0}$ and logical $\ket{1}$ qubits into the left-, $\ket{L}$ and right-handed polarisations, $\ket{R}$, and apply a Hadamard gate before and after interaction in the case of a star measurement, such that $\ket{0}\leftrightarrow\ket{+}$ and $\ket{1}\leftrightarrow\ket{-}$. Then the evolution of a photonic eigenstate and the spin state can be expressed as
\begin{equation}\label{equation:SMstate}
\begin{split}
\bigotimes_{\substack{j\in \text{star}(s) \\ \text{or }\text{plaq}(p)}}\ket{i_j}\otimes\ket{+_S}\rightarrow & \bigotimes_{j}\exp\left(\I\phi_{j,\uparrow}\right)\ket{i_j}\otimes\ket{\uparrow}/\sqrt{2}+\bigotimes_{j}\exp\left(\I\phi_{j,\downarrow}\right)\ket{i_j}\otimes\ket{\downarrow}/\sqrt{2}\\
&=\bigotimes_j \exp\left(\I\phi_{j,\uparrow}\right)\ket{i_j}\otimes\left[\ket{\uparrow}+\prod_k\exp\left[-\I\left(\phi_{k,\uparrow}-\phi_{k,\downarrow}\right)\right]\ket{\downarrow}\right]/\sqrt{2}\\
&=e^{4\I\phi_0}\bigotimes_j \exp\left(\I\tilde{\phi}\delta_{i_jL}\right)\ket{i_j}\otimes\left[\ket{\uparrow}+\prod_k\exp\left[-\I\left(\phi_{k,\uparrow}-\phi_{k,\downarrow}\right)\right]\ket{\downarrow}\right]/\sqrt{2},
\end{split}
\end{equation}
where $\ket{i_j}\in\{\ket{L},\ket{R}\}$, $j$ and $k$ index the same four photonic qubits in a plaquette or star set, $\delta_{i_j L}$ is the Kronecker delta, and $\phi_{j,\ast}$ is the phase shift resulting from the interaction between the photonic state $\ket{i_j}$ and spin state $\ket{\ast}\in\left\{\ket{\uparrow},\ket{\downarrow}\right\}$. This gives us Eq. \ref{equation:state} up to a global phase $\exp\left[4\I\phi_0\right]$. (One may also consider an initial spin state of $\ket{-_S}=\left( \ket{\uparrow}-\ket{\downarrow}\right)/\sqrt{2}$, and perform syndrome extraction in a similar manner.)

\section*{Considering boundary conditions}

\begin{figure}[]
\includegraphics[width=0.4\columnwidth]{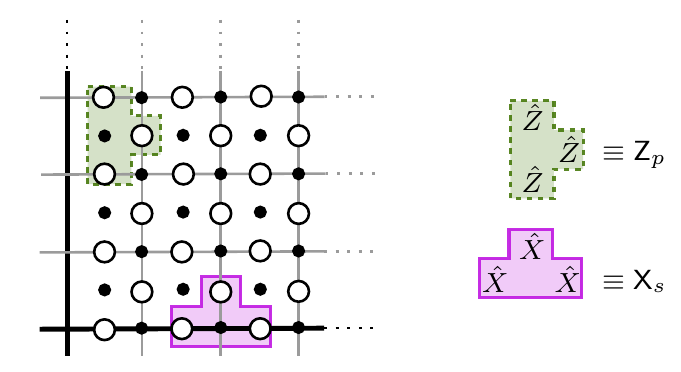}
\caption{A representation of boundary conditions in the two-dimensional surface code. The plaquette, $\mathsf{Z}_p$, and star, $\mathsf{X}_s$ measurement operators of weight three acting on the given boundaries are shown in green (dotted outline) and pink (solid outline), respectively.}
\label{fig:sandp2}
\end{figure}

So far, we have consider the toric code exhibiting only periodic boundary conditions. Different surface codes may also be comprised of various types of boundaries that result in modifications of the stabilizer operators applied on the data qubits located along these boundaries. Consider boundaries in the qubit array as shown in Fig. \ref{fig:sandp2}. The spin--photon interface can be used for these syndrome measurements as well, without requiring a change in the frequency detuning $\delta$, such that
\begin{equation}
\bigotimes_{\substack{j\in \text{star}(s) \\ \text{or }\text{plaq}(p)}}\ket{i_j}\otimes\ket{+_S}\rightarrow e^{3\I\phi_0}\bigotimes_j \exp\left(\I\tilde{\phi}\delta_{i_jL}\right)\ket{i_j}\otimes\left[\ket{\uparrow}+\prod_k\exp\left[-\I\left(\phi_{k,\uparrow}-\phi_{k,\downarrow}\right)\right]\ket{\downarrow}\right]/\sqrt{2}.\\
\end{equation}
As the cumulation of the relative phase in the electron spin state is dependent on the parity of the photonic state, where the $+1$ ($-1$) eigenstates are of even (odd) parity, the state evolves to either $\ket{L}_S=\left(\ket{\uparrow}+\I\ket{\downarrow}\right)/\sqrt{2}$ ($\ket{R}_S=\left(\ket{\uparrow}-\I\ket{\downarrow}\right)/\sqrt{2}$). The syndrome can therefore be obtained by measuring the spin in the $\hat{Y}$-basis. Any corrections to the phase shift of the photonic state are addressed in the same manner as for weight-four operators.

\section*{Swapping the assignment of the data and measure qubits}

The interaction between the photon and the spin is symmetric, meaning that the assignment of the data and measure qubits may be swapped around. We show this explicitly by starting off from, say, a horizontally polarised photon $\ket{H}=\left(\ket{L}+\ket{R}\right)/\sqrt{2}$ that then interacts consecutively with each spin state. The total state transforms to
\begin{equation}
\bigotimes_{\substack{j\in \text{star}(s) \\ \text{or }\text{plaq}(p)}}\ket{i_j}\otimes\ket{H}\rightarrow  e^{4\I\phi_0}\bigotimes_j \exp\left(\I\tilde{\phi}\delta_{i_j\uparrow}\right)\ket{i}_j\otimes\left[\ket{\L}+\prod_k\exp\left[-\I\left(\phi_{k,L}-\phi_{k,R}\right)\right]\ket{R}\right]/\sqrt{2},
\end{equation}
where now $\ket{i_j}\in\left\{\ket{\uparrow},\ket{\downarrow}\right\}$ and $\phi_{j,\ast}$ is the phase shift resulting from the interaction between the spin state $\ket{i_j}$ and spin state $\ket{\ast}\in\left\{\ket{L},\ket{R}\right\}$. The detection of a horizontally (vertically, $\ket{V}=-\I\left(\ket{L}-\ket{R}\right)/\sqrt{2}$) polarised photon would signal a syndrome of $+1$ ($-1$). In the case of a syndrome measurement at the boundaries, the photonic state evolves to $\ket{L}\pm\I\ket{R}$, which can be easily resolved into $\ket{H}$ and $\ket{V}$ by adding a polarisation-dependent $\pi/2$ phase shift post-interaction and before photon-detection.

In order to correct for possible changes in the relative phase shifts between the eigenstates making up the quiescent state, it is sufficient to simply allow a $\ket{R}$ photon to interact with each data qubit. Then
\begin{equation}\begin{split}
e^{4\I\phi_0}\bigotimes_j \exp\left(\I\tilde{\phi}\delta_{i_j\uparrow}\right)\ket{i}_j\otimes\ket{R}\rightarrow & e^{4\I\phi_0}\bigotimes_j \exp\left[\I\left(\tilde{\phi}\delta_{i_j\uparrow}+\phi_0 \delta_{i_j\uparrow}+\phi_h \delta_{i_j\downarrow}\right)\right]\ket{i}_j\otimes\ket{R}\\
& = \exp\left[4\I\left(\phi_0+\phi_h \right)\right]\bigotimes_j\ket{i}_j\otimes\ket{R},
\end{split}
\end{equation}
where $\tilde{\phi}+\phi_0=\phi_h$ and the reflection coefficient described in Eq. \ref{eq:rh} is independent of the spin orientation. This way, every eigenstate obtains an equivalent total phase shift, rendering just an overall global phase shift to the surface code.

\section*{Entanglement generation}

\begin{figure}[t]
\includegraphics[width=0.7\columnwidth]{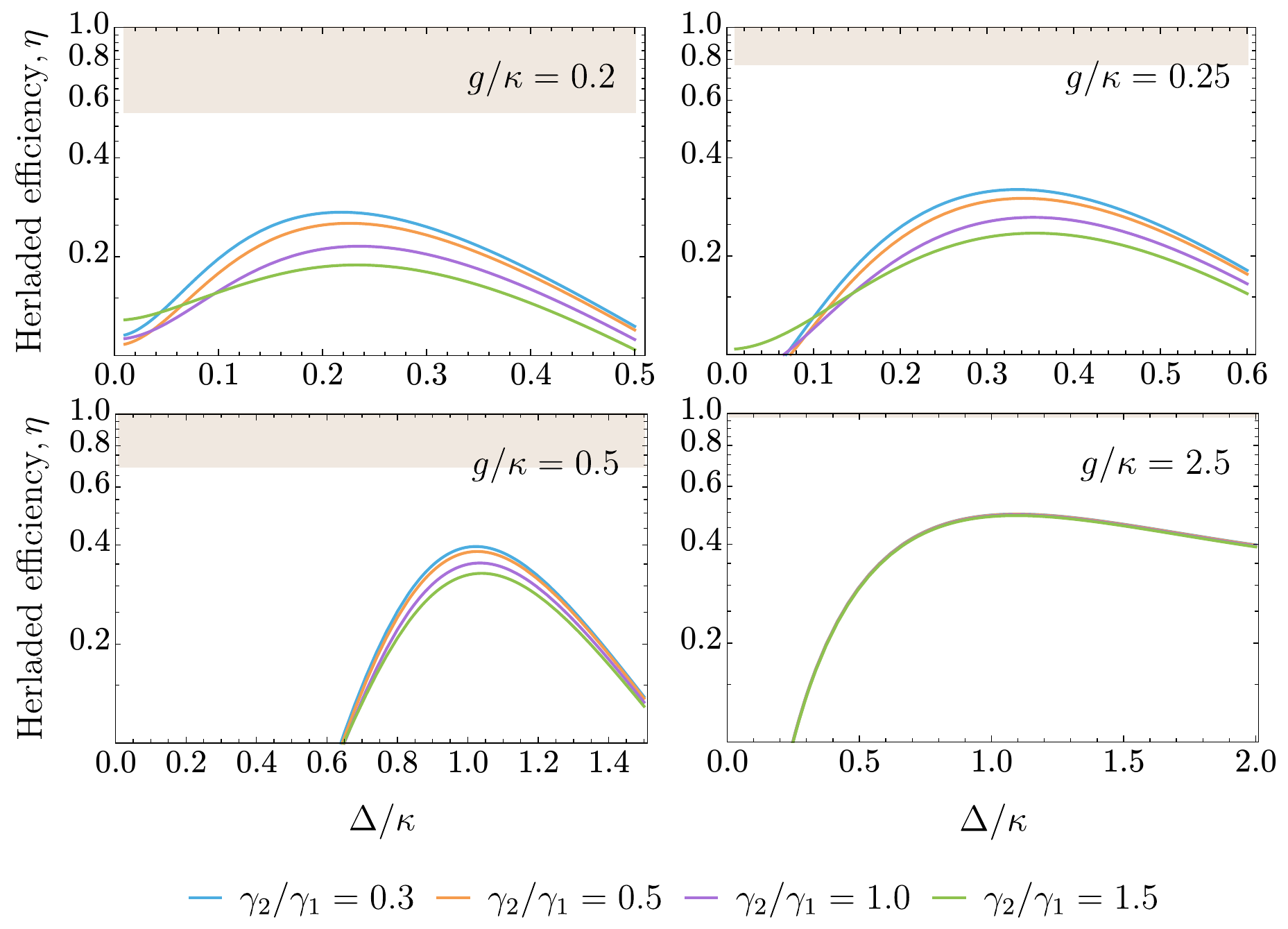}
\caption{Heralded efficiency, $\eta$, of the entanglement generation protocol as a function of $\Delta/\kappa$, where $\Delta=\omega_{X_1}-\omega_{X_2}$ is the central energy detuning, for different linewidth ratios, $\gamma_2/\gamma_1$, and coupling strengths $g/\kappa$. $\gamma_1/\kappa$ is set to 0.1. The shaded regions indicate the maximum possible efficiency of the protocol in the ideal case with identical systems.}
\label{fig:plot1}
\end{figure}

In Fig. \ref{fig:plot1} we show the heralded efficiency, $\eta$, of the entangling procedure described by Eq. \ref{eq:prob} as a function of the characteristic energy detuning, $\Delta=\omega_{X_1}-\omega_{X_2}$, for varying cavity decay rates $\kappa$ and coupling strengths $g$. (We work here both in the weak- and the strong-coupling regime to show that the entangling procedure may be employed in either.) We consider only the case of $\tilde{\phi_1}=-\tilde{\phi_2}$ due to higher probabilities of success, as phase shifts that satisfy this condition for QD--spin systems exhibiting typical variations can approach very closely $\pm\pi/2$. Looking at Eq. \ref{eq:prob}, this maximises the probability of measuring an orthogonally polarised photon. Also, variations in the QD linewidths may marginally enhance the efficiency, however the effect of this decreases as $g$ is increased. 

\begin{figure}[]
\includegraphics[width=0.7\columnwidth]{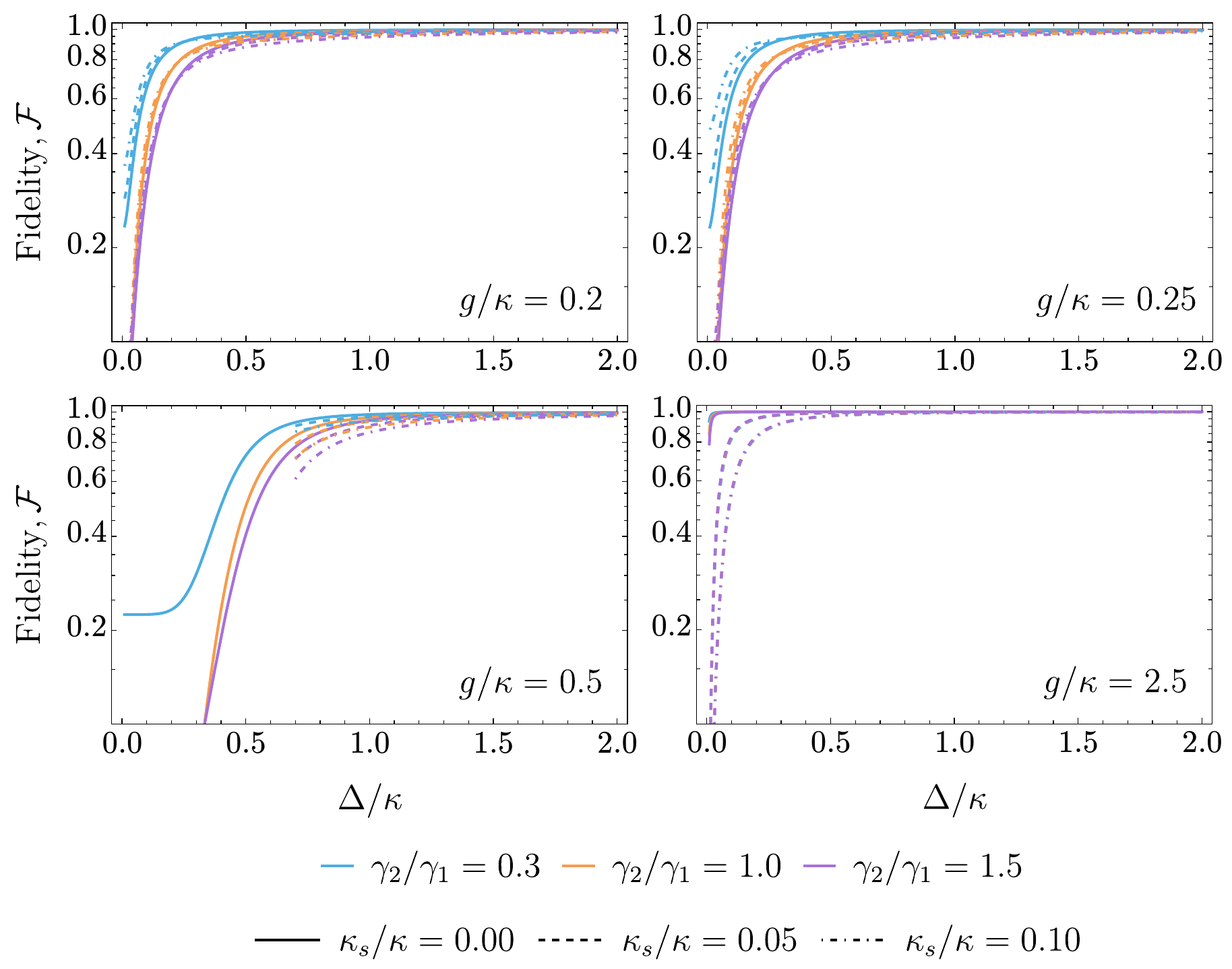}
\caption{Fidelity, $\mathcal{F}$, of the entanglement generation process as a function of $\Delta/\kappa$, where $\Delta=\omega_{X_1}-\omega_{X_2}$ is the central energy detuning. The QDs have linewidth ratios, $\gamma_2/\gamma_1$, are set to 0.3 (blue), 1.0 (orange) and 1.5 (purple); cavity loss rates, $\kappa_s/\kappa$, of 0.0 (solid), 0.2 (dashed) and 0.5 (dash-dotted); $\gamma_1/\kappa$ is set to 0.1. }
\label{fig:plot2}
\end{figure}

We also show in Fig. \ref{fig:plot2} the effect of spectral variations as well as side-cavity losses, characterised by $\kappa_s$, on the fidelity of the entangled state:
\begin{equation}
\mathcal{F}=\frac{|r_{h_1}r_{h_2}-r_{0_1}r_{0_2}|^2}{|r_{h_1}r_{h_2}-r_{0_1}r_{0_2}|^2+|r_{h_1}r_{0_2}-r_{0_1}r_{h_2}|^2},
\end{equation}
where $r_{h_i}$ and $r_{0_i}$ are the reflection coefficients for the QD-coupled and empty cavity cases for system $i$, respectively.

\end{document}